\documentclass
[aps,prl,twocolumn,showpacs,floats,floatfix,amsmath,superscriptaddress]{revtex4}
\usepackage{graphicx}
\usepackage{amsmath}
\usepackage{amsfonts}
\usepackage{amssymb}%
\usepackage{epsf}
\usepackage{rotate}

\newcommand{\be}{\begin{equation}}
\newcommand{\ee}{  \end{equation}}
\newcommand{\ba}{\begin{eqnarray}}
\newcommand{\ea}{  \end{eqnarray}}
\newcommand{\ket}[1]{\left|#1\right>}
\newcommand{\bra}[1]{\left< #1 \right|}
\newcommand{\braket}[2]{\left< #1| #2 \right>}

\begin{document}

\title{Universal decoherence induced by an environmental quantum phase transition}

\author{Fernando Mart\'{\i}n Cucchietti}
\affiliation{%
Theoretical Division, Los Alamos National Laboratory, MS B213, Los Alamos, NM 87545,USA
}%
\author{Sonia Fernandez--Vidal}
\affiliation{%
Departamento de F\'{\i}sica, Universidad Autonoma de Barcelona, 
Bellaterra, Espa\~na
}%
\author{Juan Pablo Paz}
\affiliation{%
Theoretical Division, Los Alamos National Laboratory, MS B213, Los Alamos, NM 87545,USA
}%
\affiliation{%
Departamento de F\'{\i}sica, FCEyN, UBA, Pabell\'{o}n $1$
Ciudad Universitaria, 1428 Buenos Aires, Argentina.
}%

\date{\today}%
\begin{abstract}
Decoherence induced by coupling a system with an environment may display
universal features. Here we demostrate that when the coupling to 
the system drives a quantum phase transition in the environment, the temporal 
decay of quantum coherences in the system is Gaussian with a width
independent of the system-environment coupling strength. 
The existence of this effect  opens the way for a new
type of quantum simulation algorithm, where a single qubit is used to detect 
a quantum phase transition. We discuss possible implementations
of such algorithm and we relate our results to available data on universal
decoherence in NMR echo experiments. 
\end{abstract} 


\maketitle 

The coupling between a quantum system and its environment leads to 
decoherence, the process by which quantum information is degraded. 
Decoherence plays a crucial role in the understanding of the 
quantum to classical transition \cite{deco}. It also has 
practical importance: its understanding is essential in
technologies that actively use quantum coherence, 
such as quantum information processing \cite{QIP}. 
In general, the timescale $t_{dec}$ of decoherence depends on the 
system-environment coupling strength, which we arbitrarily denote $\lambda$. 
For example, in the well studied case of quantum Brownian motion (where the 
environment consists of a large number of non--interacting harmonic oscillators), 
quantum coherence generally decays exponentially with a rate $1/t_{dec}$ 
proportional to $\lambda^2$ \cite{qbm}. In this letter we describe a class 
of systems with a drastically different behavior: Gaussian decay of coherence with
a rate independent of $\lambda$. This independence signals a universal
behavior whose study is the aim of this work.
In general, one should avoid building physical quantum information processing 
devices in presence of universal decoherence. 
However, we show that universality is a powerful property we can use to our
advantadge: by detecting decoherence in the universal
regime we can extract valuable information about the environment.

Environment-independent decoherence rates are also found in other circumstances.
For example, systems with a classically chaotic Hamiltonian 
display a ``Lyapunov regime'' where the decay is exponential and given by 
the  Lyapunov exponent of the underlying classical 
dynamics \cite{ZurekPaz,Lecho1}. These models are also often used to 
represent a {\em complex}  environment. In fact, chaoticity is the widespread 
explanation \cite{Lecho1,Lecho2} for the  perturbation-independent decay of polarization 
detected in recent NMR echo experiments \cite{Horacio} 
(where, however, a non-exponential but Gaussian decay is actually observed). 
Our findings are different from the usual exponential Lyapunov regime:  
we discuss systems where the universal (independent of $\lambda$) 
decoherence is Gaussian. In our model, the complexity and sensitivity of the
environment arise from the susceptibility of the environmental spectrum to
the system's state. The relation between our results and the experiments
of Ref. \cite{Horacio}  will also be discussed below.

Let us consider a spin $1/2$ particle (a qubit) 
coupled to an environment that is ``structurally unstable'' with respect to the system 
state (in a sense that will be made clear below). The model we discuss is a 
generalization of the one studied by Quan {\em et al} \cite{Quan}, who showed that 
an environment at the critical point of a quantum phase transition is highly efficient in producing 
decoherence. Below, we will not only generalize the results of \cite{Quan} but also
show that in these circumstances universal decoherence arises naturally.
We assume that the system 
and the environment evolve under the Hamiltonian 
\ba
{\cal H_{SE}} = {\mathcal I}_S\otimes {\cal H}_E+
\left|0\right>\left<0\right| \otimes {\cal H}_{\lambda_0} + 
\left|1\right>\left<1\right| \otimes {\cal H}_{\lambda_1}.
\label{hamiltonian}
\ea
Here, the operators ${\cal H}_E$, ${\cal H}_{\lambda_0}$ and ${\cal H}_{\lambda_1}$
act on the Hilbert space of the environment. 
If the system is in 
state $|j\rangle$ ($j=0,1$), the environment evolves with an effective Hamiltonian 
${\cal H}_j={\cal H}_E+{\cal H}_{\lambda_j}$  ($\lambda_j$ is  
the system-environment coupling strength). Considering the initial state 
$\left| \Psi_{\cal SE}(0)\right> = 
(a \left|0\right>+b \left|1\right>) \ket{{\cal E}(0)},$
the evolved reduced density matrix of the system is
\ba
\rho_{\cal S} (t) & = & {\rm Tr} _{\cal E} \left| \Psi_{\cal SE}(t)\right> 
\left< \Psi_{\cal SE}(t)\right| \nonumber \\
& = & |a|^2 \left|0\right>\left<0\right|+ a b^{*} r(t)
\left|0\right>\left<1\right| \nonumber \\ 
& + & a^{*} b r^{*}(t) \left|1\right>\left<0\right| + |b|^2 \left| 1 \right> \left< 1 \right|.
\label{reducedrho}
\ea
The off-diagonal terms of this operator are modulated by the 
decoherence factor $r(t)$: the overlap between two
states of the environment obtained by evolving the initial state
$\left| {\cal E}(0) \right>$ with two different Hamiltonians, i.e. 
$
r(t)= \left< {\cal E}(0) \right|e^{i{\cal H}_0 t} 
e^{-i{\cal H}_1 t} \left| {\cal E}(0) \right>$. 
Moreover, assuming that the initial state of the environment is the
ground state $\ket{g_0}$ of  ${\cal H}_0$ \cite{eigenstate}, the 
decoherence factor $r(t)$ is, up to an irrelevant phase factor, identical 
to the so--called survival probability amplitude
\ba
r(t)= \bra{g_0}  e^{-i{\cal H}_1 t}  \ket{g_0}. 
\label{decofactor}
\ea

Let us first analyze models where both Hamiltonians ${\cal H}_j$ ($j=0,1$)
can be diagonalized in terms of a suitable set of 
fermionic creation and annihilation operators $\gamma_k^{(j)}$: 
\be
{\cal H}_j=\sum_{k=1}^N \epsilon_k^{(j)} 
\left( \gamma_k^{(j)\dagger} \gamma_k^{(j)} - \frac{1}{2} \right). 
\label{environment}
\ee
Furthermore, we assume that the operators appearing in the 
two Hamiltonians ${\cal H}_j$ can be connected by a Bogoliubov transformation 
of the form
\be
\gamma_{k}^{(1)} = \cos(\alpha_k) \gamma_{k}^{(0)} - i 
                   \sin(\alpha_k) \gamma_{-k}^{(0)\dagger},
\label{bogoliubov}
\ee
where the angles $\alpha_k$ define the Bogoliugov coefficients. 
Notice that this expression only includes mixing between modes with opposite values 
of the index $k$. Our treatment can be extended to more complicated situations, 
but we limit first to the simplest non--trivial case, where it is possible to relate the ground 
states $\ket{g}_j$ of ${\cal H}_j$ as 
\be
\ket{g}_0 = \prod_{k>0} \left[ i \cos(\alpha_k) + \sin(\alpha_k) \gamma^{(1)\dagger}_k 
\gamma^{(1)\dagger}_{-k} \right] \ket{g}_1.
\ee

Under these assumptions the decoherence factor is 
\be
r(t)=\prod_{k>0} \left( \cos^2 (\alpha_k) e^{i t \epsilon^{(1)}_k} +
\sin^2 (\alpha_k) e^{-i t \epsilon^{(1)}_k} \right).
\label{decofactor2}
\ee
Surprisingly, $r(t)$ is completely analogous to the one found when 
studying non--interacting spin environments \cite{Cucchietti}. In 
that case, the index $k$ labels the different
environmental spins and the corresponding Bogoliubov coefficients define their  
initial states.

Under reasonable assumptions on the angles $\alpha_k$ 
and the energies $\epsilon^{(1)}_k$, 
we can go further and -- using the ideas developed
in \cite{Cucchietti} -- obtain a simple form for the temporal evolution of the overlap 
$r(t)$. To illustrate our procedure, let us analyze
first an oversimplified case: suppose that the energies of all the modes 
are the same, i.e. $\epsilon^{(1)}_k=\epsilon$. In the simplest case $\alpha_k=\pi/4$,
the overlap oscillates as $r(t)=(\cos\epsilon t)^{N/2}$. The same
result is recovered as a consecuence of the law of large numbers if the 
angles $\alpha_k$ are spread over the entire circle. In fact, 
$|r(t)|^2\simeq |\cos \epsilon t|^{N}$ if the following
Lindenberg conditions are satisfied
\ba
&{1\over N}&\sum_k\cos^2\alpha_k \simeq 1/2 \nonumber \\ 
s_N^2&=&\sum_k \sin^2 2\alpha_k \left(\epsilon^{(1)}_k\right)^2
\gg \epsilon^2.\label{Lindenberg}
\ea
The first condition is satisfied when the angles 
are randomly distributed.
The second one imposes a finite variance for the ``quantum walk'' 
in which a step of length $+\epsilon_k$ ($-\epsilon_k$) is taken with 
probability $\cos^2\alpha_k$ ($\sin^2\alpha_k$). 
When $\epsilon^{(1)}_k=\epsilon$, the condition takes the form $s_N^2 \gg 1$, and
it is met when there is a sufficiently large number of modes for which 
$\sin 2 \alpha_k$ does not vanish. 

A more realistic situation is when the energies $\epsilon^{(1)}_k$ take values in a 
given spectral band. When the energies are distributed with a vanishing 
mean value, the decay of $r(t)$ is Gaussian with a width given by $s_N^2$ defined 
in (\ref{Lindenberg}) 
\cite{Cucchietti}. 
Consider the more general case 
where the energies are distributed about an arbitrary mean value, 
i.e. $\epsilon_k^{(1)}=\epsilon + \delta_k$ (where $\delta_k$ has zero mean). 
We now define the dispersion $\tilde{s}_N^2$ as the cumulative variance 
of the fluctuations of the energy, i.e. $\tilde{s}_N^2=\sum_k \sin^2 2\alpha_k\ \delta_k^2$.
We find that, in general, when conditions (\ref{Lindenberg}) hold (replacing $s_N^2$ by 
$\tilde{s}_N^2$), $r(t)$ is described by a Gaussian envelope 
modulating an oscillating term,
\be
|r(t)|^2=\exp(- \tilde{s}_N^2\ t^2) |\cos(\epsilon t)|^{N/2}.\label{roft}
\ee

In general, when the operators $\gamma^{(0)}_k$ and $\gamma^{(1)}_k$ are 
similar, the angles $\alpha_k$ are small and (\ref{Lindenberg}) do not hold:
there is almost no decoherence. However, a drastic difference in the nature of the
eigenstates of ${\cal H}_0$ and ${\cal H}_1$ can only be accounted for with 
$\alpha_k$ varying in the full range $[0,2\pi)$. This occurs 
when the environment suffers a quantum phase transition when $\lambda$ is varied. 
Thus, denoting $\lambda_c$ the critical point
of the transition, for $\lambda_0 \ll \lambda_c \ll \lambda_1$ 
we expect the decoherence factor to behave as indicated in (\ref{roft}). 
In many cases, $\tilde{s}_N^2$ is only given by the properties of the environment 
Hamiltonian, and thus the decay of $r(t)$ becomes {\em universal} (independent of $\lambda$).

An important model encompassed by assumptions (\ref{environment}) and
(\ref{bogoliubov}) is an Ising chain transversely coupled to a central spin 
\cite{Quan} (which plays the role of the system). In this case 
\be
{\cal H}_j=-J\bigl(\sum_{i=1}^N \sigma_i^z\sigma_{i+1}^z
-\lambda_j\sum_{i=1}^N\sigma_i^x\bigr).
\label{HIsing}
\ee
The Bogoliubov coefficients and the energies are \cite{Sachdev},
\begin{eqnarray}
\epsilon_k^{(j)}&=&2J \sqrt{1+\lambda_j^2-2\lambda_j\cos(2\pi k/N)}\\
2\alpha_k&=&(\theta_k(\lambda_1)-\theta_k(\lambda_0)),\label{epsilonalphaising}
\end{eqnarray}
where the angles $\theta_k(\lambda)$ are defined from
$\tan(\theta_k)=\sin(2\pi k/N) / (\lambda - \cos(2\pi k/N))$. In this model $\lambda_c=1$.

When $\lambda_1\gg 1$ and $\lambda_0 < 1$, the angles 
$\alpha_k(\lambda)\approx \pi k/N$ and the Bogoliubov coefficients
satisfy conditions (\ref{Lindenberg}). 
Moreover, the energies $\epsilon_k^{(1)}$ are distributed 
between $|\lambda_1-1|$ and $|\lambda_1+1|$, which gives $\tilde{s}^2_N\approx N$. 
Therefore, the width of the Gaussian 
envelope is {\it independent of $\lambda_1$}.

\begin{figure}[tb]
\centering \leavevmode
\epsfxsize 3.2in
\epsfbox{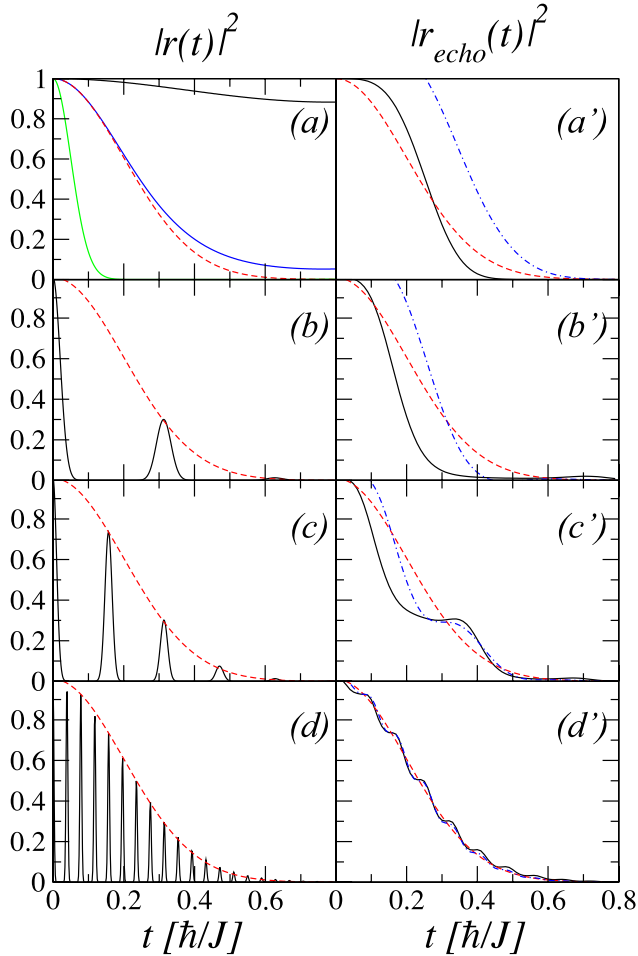}
\caption{Decoherence factor for the Ising model Hamiltonian as the environment with 
50 spins for (a) $\lambda_1=2$,
(b) $\lambda_1=5$, (c) $\lambda_1=10$, and (d) $\lambda_1=40$. 
In (a) the two top curves included for comparison have $\lambda_1=0.1$ 
and $\lambda_1=0.5$. In all plots $\lambda_0=0$, and the dashed line
is the predicted universal Gaussian envelope. On the right, the same 
values of $\lambda_1$ for $r_{echo}(t)$, which eliminates the 
$\lambda_1$ dependent oscillations.}
\label{Figure1}
\end{figure}

In Fig. 1 we display $r(t)$ for the case $\lambda_0=0$, showing the accuracy of 
Eq. (\ref{roft}). The {\it universality} of the envelope is a clear 
indication of the quantum phase transition. However, the oscillations 
(whose frequency depends on $\lambda_1$) are not universal. 
Yet, it is possible to eliminate them by performing a spin-echo experiment:
first, evolve the system coupled to the Ising chain environment
for a time $t$. At this time, flip the environmental spins in the $x$ direction 
(e.g. with an rf-pulse that applies a $\pi$-rotation around the $z$-axis). 
Finally, evolve for another time $t$. 
The total evolution of the environment can be described by using 
the Hamiltonian ${\cal H}_1={\cal H}_E+{\cal H}_{\lambda_1}$ from time $0$ to $t$, 
and ${\cal H}_{-1}={\cal H}_E-{\cal H}_{\lambda_1}$ 
from time $t$ to $2t$. Thus, in this echo experiment the 
decoherence factor is given by
\be
r_{echo}(2t)= \bra{g}_0  e^{-i{\cal H}_{-1} t} e^{-i{\cal H}_1 t}  \ket{g}_0. 
\label{decofactorecho}
\ee
This overlap is simply computed using the Bogoliubov transformation
that connect the modes diagonalizing the Hamiltonians ${\cal H}_{-1}$ and ${\cal H}_0$. 
If we denote  $\gamma_k^{(-1)}$ the modes of ${\cal H}_{-1}$, 
the Bogoliubov coefficients associated with the corresponding angles $\tilde \alpha_k$ 
are such that $\gamma_{k}^{(-1)} = \cos(\tilde\alpha_k) \gamma_{k}^{(0)} - i 
                    \sin(\tilde\alpha_k) \gamma_{-k}^{(0)\dagger}$. The analytic form for the overlap $r_{echo}(t)$ is simplified introducing the sum and difference of the 
energies, $\epsilon_k^{(\pm)}=\epsilon_k^{(1)}\pm
\epsilon_k^{(-1)}$, and the  
Bogoliubov angles, $\alpha_k^{(\pm)}=\tilde\alpha_k \pm \alpha_k$. We obtain
\ba
r_{echo}(2t)&=&\prod_{k>0}\bigl[
\cos\epsilon_k^{(+)}t\ \cos^2\alpha_k^{(-)} 
+ \cos\epsilon_k^{(-)}t\ \sin^2\alpha_k^{(-)}\nonumber\\ 
&+& i \sin\epsilon_k^{(+)}t\ \cos\alpha_k^{(-)}\cos\alpha_k^{(+)}\nonumber \\
&+& i \sin\epsilon_k^{(-)}t\ \sin\alpha_k^{(-)}\sin\alpha_k^{(+)} \bigr].
\label{recho1}
\ea
For the case of the Ising model the expression 
can be evaluated explicitely. In 
the limit of large values of $\lambda$, one can obtain an 
approximate behavior using 
similar arguments as above \cite{remark}. Thus, the dominant contribution to 
the echo--overlap is 
\be
r_{echo}(t)\approx \exp(-\tilde{s}_N^2t^2) \left(1 - \frac{K(t)}{\lambda} \sin(\lambda t)\right),
\ee
where $K(t)=2 \sum_k \sin (\epsilon_k^{(-)} t) \cos (2\pi k/N) \sin^2(2\pi k/N) $.
In Fig. 1 we show how the accuracy of this expression increases with $\lambda$.

To test the generality of our results against the restrictiveness and uncontrollability 
of assumptions (\ref{environment}) and (\ref{bogoliubov}), we study a system 
in the opposite end of the spectrum: the Bose-Hubbard model (BHM) \cite{Sachdev},
with Hamiltonian
\be
{\cal H}_{BH}=-g \sum_{<i,j>} a^\dagger_i a_j + u \sum_n a^\dagger_n a_n(a^\dagger_n a_n-1).
\ee
Here $a_n$ are boson anihilation operators in site $n$ of a discrete 
lattice. For $g\gg u$, the system behaves as a superfluid of non--interacting 
particles. In the opposite regime, $u \gg g$, 
the interaction term dominates and the ground state is Mott-insulator like.
This model cannot be cast in terms of fermionic operators as in (\ref{environment}), 
in fact, no analytic solution is known. Furthermore, the bosonic 
nature of the particles also conflicts
with (\ref{bogoliubov}). The BHM has practical relevance because it 
can be experimentally simulated using cold neutral
atoms in an optical lattice \cite{Bloch}.
We calculate $r(t)$ numerically 
for a spin $1/2$ coupled to the hopping term of the BH Hamiltonian, that is,
we take $g\equiv \lambda$. In Fig. 2 we show the decoherence factor for several values of 
$\lambda$ for a BHM with a fixed number of bosons. The same overall behavior of the 
Ising chain is observed: a universal Gaussian envelope (independent of $\lambda$) modulating
an oscillation with frequency proportional to $\lambda$. The very different nature of the BHM
hints at a more general validity of our results.

\begin{figure}[tb]
\centering \leavevmode
\epsfxsize 3.2in
\epsfbox{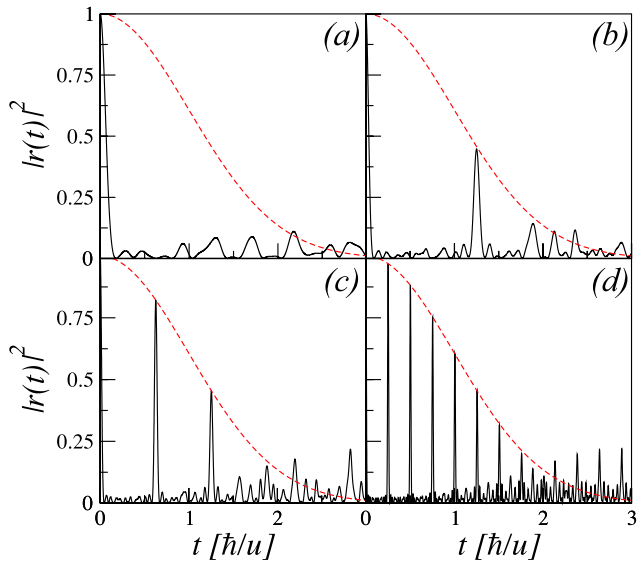}
\caption{Decoherence factor for the Bose-Hubbard model Hamiltonian as the environment, with 
6 particles in a lattice of 6 sites, for (a) $J=5$,
(b) $J=10$, (c) $J=20$, and (d) $J=50$. In all plots the dashed line
is the universal Gaussian envelope whose width is numerically obtained.}
\label{Figure2}
\end{figure}

A Gaussian decay of coherence with a rate independent of the 
coupling to the environment was indeed observed in NMR polarization echo 
experiments \cite{Horacio}. 
Arguing on the complexity of the experimental many-body system,
these results have been related to the environment-independent 
decoherence predicted in classically chaotic Hamiltonians \cite{ZurekPaz,Lecho1,Lecho2}.
The experimental situation is quite different from 
the one we considered here: the decoherence factor is measured after 
an echo created by a change of sign of the environment Hamiltonian, 
and not the system-bath interaction.
Our model points to a different way of introducing complexity
and sensitivity in the environment: a quantum phase transition. 
Further research using this approach might explore more realistic models 
that account for all the details of the experiments.

The universal decoherence regime of this work can also be understood
using analogies to the regime of strong perturbations of the survival 
probability, Eq. (\ref{decofactor}). 
Indeed,  $r(t)$ is the Fourier transform of 
the strength function or local density of states (LDOS), 
$L(E)=\sum_n |\braket{g_0}{\phi_n}|^2 \delta(E-E_n)$, where 
$\ket{\phi_n}$ are the eigenvectors of ${\cal H}_1$ and $E_n$ its
eigenenergies. In typical LDOS studies, ${\cal H}_1$ differs from
${\cal H}_0$ by a perturbation.
In complex systems (e.g. random
matrices, or classically chaotic Hamiltonians) for sufficiently
strong perturbations $\ket{g_0}$ is a random
superposition of the $\ket{\phi_n}$ states. Therefore, 
the LDOS becomes independent of the perturbation:
it equals the full density of states of ${\cal H}_1$. In our model, the saturation
of the LDOS when ${\cal H}_0$ and ${\cal H}_1$ are
on both sides of the quantum phase transition occurs because of the
radically different nature of the eigenstates.
In contrast to our results, Refs. \cite{Heller} have found that complex systems give an 
LDOS with a Lorentzian shape, leading to an exponential decay
of $r(t)$.

Universal decoherence can be harmful for quantum information applications.
However, it can be a useful tool to extract 
information about a critical system, e.g.  its spectral structure or the critical point of 
its quantum phase transition. The latter example can
be thought of as a ``critical point finding'' algorithm in a 
one-qubit quantum computer: in systems where the spectrum is not shifted by the 
coupling (which gives the oscillatory $cos(\lambda t)^N$ term), 
the critical point can be simply obtained as the 
$\lambda$ value for which one observes the onset of universality. Otherwise,
the oscillation term obscures the critical point. 
In these cases one can instead couple the system weakly to
the environment, and drive the transition with an external parameter (as in
Ref. \cite{Quan}). The critical point is then signaled by the $\lambda$ value
for which there is a maximum decoherence decay. 
A demonstration of this algorithm can be
performed in an NMR setting simulating the Ising Hamiltonian studied above \cite{Raymond}.

We have shown that when the coupling to the system drives a quantum phase transition
in the environment, the decoherence factor decays as a Gaussian with 
an environment-independent width. 
We showed numerically that our findings are more general than what can be expected
from the analytical approximations we used. Our results could lead to an alternative
interpretation of hitherto unexplained NMR experimental results on environment
independent decoherence rates. Finally, we discussed how the universal 
behavior of the decoherence factor can be used to study critical systems 
in a novel simulation algorithm for one-qubit quantum computers.
We acknowledge fruitful discussions with W.H. Zurek.


\begin{thebibliography}{99}
\bibitem{remark} For $\lambda \ll 0$ the solution of $\tan(\theta_k)\sim 0$ has to be taken 
from the second branch of the tangent, $\theta_k\sim \pi$, to satisfy the symmetries of the
Ising Hamiltonian (\ref{HIsing}), giving $\alpha_k^{(-)}\approx \pi/2$.
\end{thebibliography}
\end{document}